\documentclass[showpacs,aps,prl,twocolumn,superscriptaddress]{revtex4}

\usepackage{graphicx}% Include figure files

\begin{document}

%\preprint{}

\title{Imaging phase separation near the Mott boundary in the correlated organic superconductors $\kappa$-(BEDT-TTF)$_{2}$$X$}


\author{T. Sasaki}
%\email{takahiko@imr.tohoku.ac.jp}
\author{N. Yoneyama}
\author{N. Kobayashi}
%\thanks{}
\affiliation{Institute for Materials Research, Tohoku University, Katahira 2-1-1, Sendai 980-8577, Japan}

\author{Y. Ikemoto}
\author{H. Kimura}
\affiliation{SPring-8, Japan Science Radiation Research Institute, Mikazuki, Hyogo 679-5198, Japan}
\date{\today}

\begin{abstract}

Electronic phase separation consisting of the metallic and insulating domains with 50 -- 100 $\mu$m in diameter is found in the organic Mott system $\kappa$-[($h$8-BEDT-TTF)$_{1-x}$($d$8-BEDT-TTF)$_{x}$]$_{2}$Cu[N(CN)$_{2}$]Br by means of scanning micro-region infrared spectroscopy using the synchrotron radiation.  
The phase separation appears below the critical end temperature 35 -- 40 K of the first order Mott transition.  
The observation of the macroscopic size of the domains indicates a different class of the intrinsic electronic inhomogeneity from the nano-scale one reported in the inorganic Mott systems such as High-$T_{c}$ copper and manganese oxides.

\end{abstract}

%\keywords{}

\maketitle


%Introduction

Microscopic spatially inhomogeneous electronic states have been recently observed in many kinds of correlated electron systems.  
Nano-scale spatial variation of the superconducting gap has been revealed in the superconducting state of Bi$_{2}$Sr$_{2}$CaCu$_{2}$O$_{8+\delta}$ by the scanning tunnelling spectroscopy and microscopy \cite{Lang}.  
In the normal state, charge carriers doped into antiferromagnetic insulators tend to group into some regions of the sample in the form of stripes in some copper oxides \cite{Tranquada}.
Meanwhile a different kind of the microscopic phase separation takes place in half-doped manganese oxides \cite{Fath}.  
Small variation from half doping causes phase segregation of electron-rich ferromagnetic and electron-poor antiferromagnetic domains with submicron size within the charge ordered phase.  
In the system with Mott transition, the nano-scale electronic inhomogeneity with preferred orientation has been found in slightly doped Mott insulator Ca$_{2-x}$Na$_{x}$CuO$_{2}$Cl$_{2}$ \cite{Kohsaka}.  
NiS$_{2-x}$Se$_{x}$ pyrite which is the band width controlled Mott system has shown also microscopic electronic inhomogeneity at the critical vicinity of the metal-insulator transition \cite{Hanaguri}.
These microscopic spatial electronic inhomogeneities seem to be intrinsic nature near the criticality of changes in charge, spin, orbital, and lattice degrees of freedom in the correlated electron system.  

Organic charge transfer salts based on the donor molecule bis(ethylenedithio)-tetrathiafulvalene, abbreviated BEDT-TTF or ET, have been recognized as one of the highly correlated electron system \cite{Michael}.  
Among them, $\kappa$-(ET)$_{2}$$X$ with $X =$ Cu(NCS)$_{2}$, Cu[N(CN)$_{2}$]$Y$ ($Y =$ Br and Cl), etc. have attracted considerable attention from the point of view of the strongly correlated quasi two dimensional electron system because the strong dimer structure consisting of two ET molecules makes the conduction band effectively half-filling \cite{Kanoda,Kino,McKenzie}.  
The unconventional metallic, antiferromagnetic insulating and superconducting phases appear next to one another in the phase diagram \cite{Kanoda,Sasaki1,Limelette}.
The transitions among these phases are controlled by the applied pressure \cite{Limelette} and slight chemical substitution of the donor and anion molecules \cite{Kawamoto}, which must change the conduction band width $W$ with respect to the effective Coulomb repulsion $U$ between two electrons on a dimer.  
Thus the $\kappa$-(ET)$_{2}$$X$ family has been considered to be the band width controlled Mott system in comparison with the filling controlled one in the inorganic perovskites such as High-$T_{\rm c}$ copper oxides.

Recently inhomogeneous electronic states have been suggested in the $^{13}$C-NMR experiments near the first order metal-insulator transition in the artificially band width controlled $\kappa$-(ET)$_{2}$Cu[N(CN)$_{2}$]Br \cite{Miyagawa}.  
Below characteristic temperature $T^{*}$ where the incoherent bad metallic state changes to the coherent good metal at lower temperature \cite{Sasaki2}, $^{13}$C-NMR lines fall into two groups indicating the metallic and antiferromagnetic insulating nature.  
The results imply that two phases coexist spatially and statically.  
Subsequent transport experiments have suggested also such coexistence of two phases at low temperature \cite{Taniguchi}.  
Although it has been demonstrated that an inhomogeneous electronic state is realized near the first order transition, the detail of the morphology, spatial distribution, size of domains and stability of the inhomogeneity have not been clarified yet.  
It is very important to obtain the real space information which can give us a clue to know either similar nano-scale electronic inhomogeneity is realized by the exotic mechanism based on the strong correlation effect or macroscopic phase separation occurs due to local potential modulation near the first order transition.  

In this letter we present the real space imaging of the electronic phase separation in the partly substituted $\kappa$-[($h$8-ET)$_{1-x}$($d$8-ET)$_{x}$]$_{2}$Cu[N(CN)$_{2}$]Br, where $h$8-ET and $d$8-ET denote fully hydrogenated and deuterated ET molecules, respectively.  
Scanning micro region infrared reflectance spectroscopy (SMIS) using the synchrotron radiation (SR) is applied to make the two dimensional map of the local electronic state.  
The results indicate that the macroscopic electronic phase separation takes place near the first order metal-insulator transition, which is different from the nano-scale electronic inhomogeneity reported so far in the inorganic correlated electron system.  

%Experiments

Single crystals of $\kappa$-[($h$8-ET)$_{1-x}$($d$8-ET)$_{x}$]$_{2}$Cu[N(CN)$_{2}$]Br partly substituted by deuterated ET molecule were grown by the standard electrochemical oxidation method.  
The substitution $x$ denotes the nominal mole ratio to the fully deuterated ET molecule in the crystallization.
We checked the actual substitution with respect to the nominal value $x$ by measuring the intensity of the molecular vibrational mode of the terminal ethylene groups.
The substitution dependence of the macroscopic phase diagram and the superconducting properties have been examined \cite{Yoneyama1}.  
The full volume of the superconductivity has been observed in the range of $x = 0 - 0.5$ when the samples are cooled slowly.  
Above $x = 0.5$, however, the superconducting volume fraction decreases and becomes about a few ten vol\% at $x = 1$, which value strongly depends on the cooling condition.  

SMIS measurements were performed using SR at BL43IR in SPring-8 \cite{Hiroaki}.  
The polarized reflectance spectra were measured on the $c$-$a$ plane along $E \parallel a$-axis and $E \parallel c$-axis with a Fourier transform spectrometer and a polarizer in the mid-infrared (IR) range by use of a mercury-cadmium-telluride detector at 77 K.  
An IR microscope with the controlled precision $x$-$y$ stage and high intensity of SR light enable us to obtain the two dimensional reflectance spectrum map with the spatial resolution of $\sim$ 10 $\mu$m \cite{Kimura}.
The sample was fixed by the conductive carbon paste on the sample holder with a gold mirror which was placed at the cold head of the helium flow type refrigerator.  
We gave careful consideration of less stress and good thermal contact to the crystals in the sample setting.  
The reflectivity was obtained by comparison with the gold mirror at each temperature measured.

%Results and Discussion

\begin{figure}
\includegraphics[viewport=3.5cm 3.3cm 17cm 21.5cm,clip,width=0.8\linewidth]{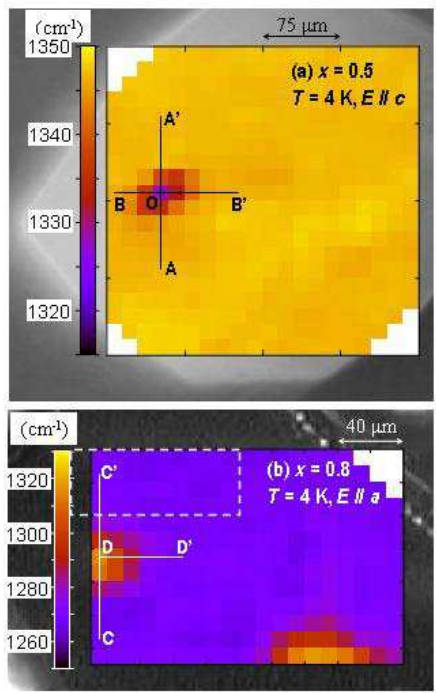}

\caption{(Color online) Two dimensional contour map of the peak frequency of the $\nu_{3}(a_{g})$ mode on the crystal surface of $\kappa$-[($h$8-ET)$_{1-x}$($d$8-ET)$_{x}$]$_{2}$Cu[N(CN)$_{2}$]Br with (a) $x =$ 0.5 ($E \parallel c$) and (b) $x =$ 0.8 ($E \parallel a$) at 4 K.  Bright color indicates the metallic nature and dark color does the insulating one. The lines AOA', BOB', COC' and DD' indicate the scanning lines shown in Figs. 2(a) and 2(b). In the region framed by dashed lines, more fine mapping was performed. }
\end{figure}

In order to make the real space image of the electronic states on the crystal surface by SMIS measurements, we use the shift of the frequency $\omega_{3}$ of a molecular vibration mode $\nu_{3}(a_{g})$.  
The specific $\nu_{3}(a_{g})$ mode, which is a symmetric stretching mode of the central double bonded carbon atoms of ET molecule, has been found to be very sensitive to difference between metallic and insulating states due to the large electron-molecular vibration coupling \cite{Sasaki2}. 
The peak of the $\nu_{3}(a_{g})$ mode should shift to lower frequency in sharper shape in the insulating state at low temperature, while it shows opposite feature in the metallic state.  
Figure 1 shows the two dimensional contour map of the reflectivity peak frequency $\omega_{3}$ of $\nu_{3}(a_{g})$ at 4 K in $\kappa$-[($h$8-ET)$_{1-x}$($d$8-ET)$_{x}$]$_{2}$Cu[N(CN)$_{2}$]Br of (a) $x =$ 0.5 and (b) $x =$ 0.8. 
The polarized reflectance spectra ($E \parallel a$-axis and $E \parallel c$-axis ) in the micro region of $\sim$ 10 $\mu$m$\phi$ are taken with a step of 15 and 10 $\mu$m for $x =$ 0.5 and 0.8 samples, respectively.  
The typical reflectivity spectra of $\nu_{3}(a_{g})$ in $E \parallel$ $c$-axis are shown in Fig. 2(c), which are taken at O point in the dark color region and B' point in the bright color region of $x =$ 0.5 sample. 
Bright region indicates the higher frequency of $\omega_{3}$ which demonstrates the metallic feature \cite{SC}.  
It is noted that the different absolute values of $\omega_{3}$ observed in the metallic (or insulating) region of $x =$ 0.5 and 0.8 samples are caused by the the polarization dependence of the $\nu_{3}(a_{g})$ mode \cite{Sasaki2}.    
Some domains with lower and higher $\omega_{3}$ can be found in the major metallic and insulating regions in $x =$ 0.5 and 0.8, respectively.  
In $x =$ 0.5, the bright region is dominant almost all over the surface.  
In contrast the dark region is dominant in $x =$ 0.8.  
The structure and position of the domain are found to be stable on time, which can be confirmed by the mapping time ($\sim$ 6 - 8 hours / one map) in SMIS measurement.  
The domains seem not to be located at particular position such as the sample edge, step and scratch of the surface and so on. 
The shape is almost circle and no specific orientation with respect to the crystal axes is observed.  

\begin{figure}
\includegraphics[viewport=3cm 3cm 18cm 23.7cm,clip,width=0.8\linewidth]{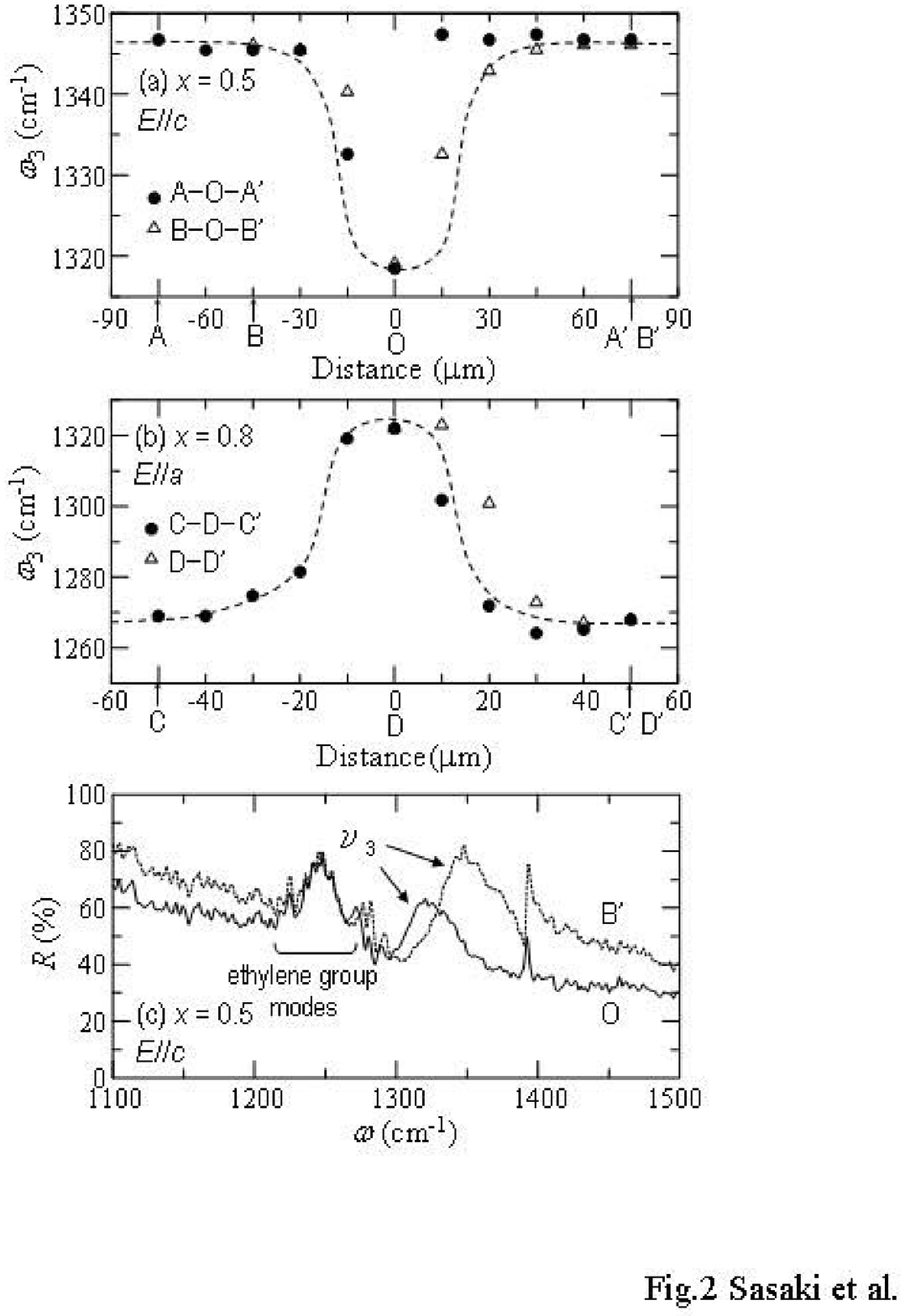}
\caption{Space variation of the peak frequency $\omega_{3}$ of the $\nu_{3}(a_{g})$ mode in $\kappa$-[($h$8-ET)$_{1-x}$($d$8-ET)$_{x}$]$_{2}$Cu[N(CN)$_{2}$]Br with (a) $x =$ 0.5 and (b) $x =$ 0.8.  The dashed curves are guides for eyes. The locations O, A, B, $\cdots$ correspond to the points indicated in Fig. 1.  (c) The reflectivity spectra at O and B' points.}
\end{figure}

Figures 2(a) and 2(b) show the scanning position dependence of $\omega_{3}$ around the insulating and metallic domains of $x =$ 0.5 and 0.8 samples, respectively.  
Each domain has a size of around $\sim$ 50 $\times$ 50 $\mu$m$^{2}$.  
The total insulating area is very roughly estimated to be $\sim$ 1/40 and $\sim$ 9/10 of the measured area for $x =$ 0.5 and 0.8 samples, respectively.   
In addition the $x =$ 0.75 sample which is not shown here has the insulating fraction of about 5/6.
This insulating fraction reflects the bulk properties obtained by the magnetization measurements \cite{Yoneyama1}.  
The superconducting volume fraction is almost the same with the sample volume from $x =$ 0 to  $\sim$ 0.5 in slow cooling condition.  
But the fraction starts to decrease with increasing $x$ from $\sim$ 0.5 to higher value.
The supercondcuting volume fraction around $x =$ 0.75 -- 0.8 is expected to be in the range of a few ten vol\% in slow cooling and a few vol\% in fast cooling.  
Considering the variation of the fraction with the cooling condition and difference of the evaluation methods, consistency between the bulk magnetization and the present measurements on the metal-insulator fraction is reasonably well.  

We have not detected the smaller domain in the present SMIS measurements. 
Inside the rectangular region framed by dashed lines in Fig. 1(b), more fine mapping (3 $\mu$m step) was performed but the spectra were the same with each other.
This does not exclude the possibility of the nano-scale inhomogeneity inside each scanning spot because the obtained spectrum may result in the average of nano-scale inhomogeneity in the measured spot. 
But close agreement between the magnetization and present results suggests that the phase separation occurs on macroscopic scale.

Possible chemical inhomogeneity to the origin of the domain structure such as segregation of deuterated ET can be excluded by checking the molecular vibration mode of terminal ethylenes of ET at each scanning point.  
The vibration modes of the ethylene groups and the deuterated one appear at different frequencies around 1250 cm$^{-1}$ and 1050 cm$^{-1}$, respectively.  
As can be seen in Fig. 2(c), almost same structure and intensity of the ethylene mode are observed at both O and B' points, which demonstrates the same substitution ratio of the deuterated ET molecule at the insulating and metallic regions.  
Therefore the present finding of the domain structure strongly indicates that the {\it electronic phase separation appears in the macroscopic size} due to the strong correlation effect near the Mott metal-insulator transition.  

\begin{figure}
\includegraphics[viewport=3.5cm 1.5cm 16.5cm 24.7cm,clip,width=0.8\linewidth]{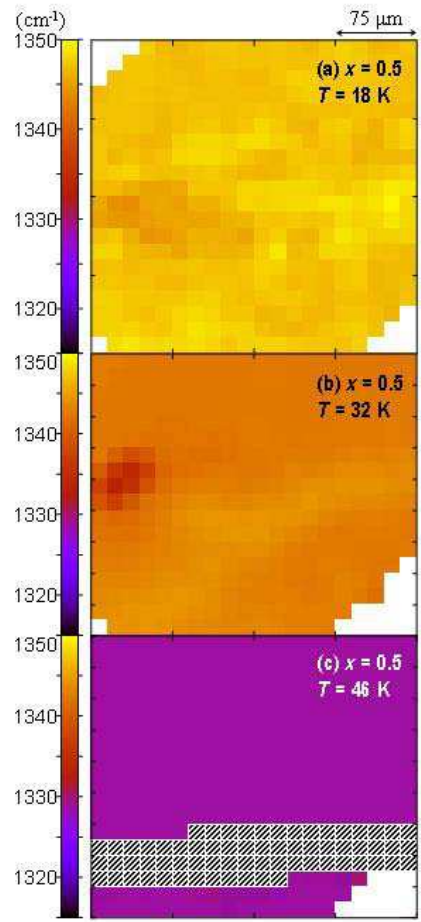}
\caption{(Color online) Temperature variation of the two dimensional contour map of the peak frequency of the $\nu_{3}(a_{g})$ mode in $E \parallel$ $c$-axis in $x =$ 0.5 sample.  Imaging region is the same with that in Fig. 1(a).   Picture elements are accidentally missing in the lower part of the map at 46 K \cite{beam}.}
\end{figure}

In order to know the correlation between the formation of the phase separation and the electronic phase diagram, the temperature variation of the phase separation was measured.  
Figure 3 shows the two dimensional contour maps on the reflectivity peak frequency $\omega_{3}$ of the $\nu_{3}(a_{g})$ mode in $E \parallel$ $c$-axis at 18, 32 and 46 K in $x =$ 0.5 sample \cite{beam}.  
The scanning region is almost the same with the map at 4 K in Fig. 2(a).
The measurements were performed from low (4 K) to high (46 K) temperature in sequence.  
The insulating domain does not change the position and size at higher temperature of 18 and 32 K.  
But the domain looks likely to disappear at 46 K. 
This temperature corresponds to the critical end point $T_{\rm cr} \simeq$ 40 K of the Mott first order metal-insulator transition \cite{Sasaki1,Limelette,Sasaki2}. 
From $T_{\rm cr}$ to both weak and strong correlation sides in the phase diagram, the $T^{*}$ line and the bad metal - insulator line $T_{\rm ins}$ are elongated.
In temperature above $T_{\rm cr}$, $T^{*}$, and $T_{\rm ins}$, the half-filling bad metallic state exists in wide range of the correlation strength which can be tuned by pressure and substitution of anion. 
The critical point in the present system has been considered to be located around $x =$ 0.5, where the phase separation may start to appear in larger $x$ value than 0.5 \cite{Yoneyama1}.
In the weak correlation side from $T_{\rm cr}$, which corresponds to the present system with $x \leq$ $\sim$ 0.5, the bad metal changes to a correlated good metal through $T^{*}$ and then becomes superconducting \cite{Limelette,Sasaki2}.  
In the strong correlation side, the bad metal develops into a Mott insulator through $T_{\rm ins}$ and then becomes an antiferromagnetic Mott insulator at $T_{\rm N}$ \cite{Limelette,Sasaki2}.  
Disappearance of the domain structure at 46 K can be explained by no multi-electronic states competing with each other above $T_{\rm cr}$, $T^{*}$, and $T_{\rm ins}$.  
On one hand below $T_{\rm cr}$, the phase separation occurs near the boundary of the first-order transition between the Mott insulator and the correlated good metal.  

In the phase separation the multiple potential minima of the free energy is required in general and the domain grows from a nucleation point which should be specific in the free energy variation in space.  
The possible origin of the space variation in the free energy is a glassy conformational order - disorder of the terminal ethylene groups of ET molecules \cite{Michael,Mueller}.  
The ethylene groups have been known to have the conformational disorder which is frozen by cooling through a temperature $T_{\rm glass} \simeq$ 80 K.  
The degree of disorder depends on the cooling rate; cooling faster introduces larger number of disorders.
Such disorder has been considered to modulate the electronic states locally \cite{Yoneyama2,Yoneyama3}.  
The slight modulation of a potential energy in space may become a nucleation point of the domain growth in the phase separation.  
In order to make it clear that the mechanism of the phase separation and the process of the domain formation, space imaging technique with the molecular resolution must be developed.  

In conclusion, the experimental evidence of the electronic phase separation is obtained by using the the real space imaging technique on the single crystal surface of the organic Mott system $\kappa$-[($h$8-ET)$_{1-x}$($d$8-ET)$_{x}$]$_{2}$Cu[N(CN)$_{2}$]Br.  
SMIS measurements using SR enable us to show the macroscopic size of the domain structure of the insulating and metallic regions.  
The observation of the micro-meter scale phase separation is different from the recent findings of the nano-scale electronic inhomogeneity in the strongly correlated inorganic system.  
The origin of the phase separation may be the combination of the strong electronic correlation near the Mott transition and the characteristic structural disorder inside the ET  molecules.

%acknowledgments

We are grateful to T. Hirono and T. Kawase for their technical supports.
SR experiments were performed at SPring-8 with the approval of JASRI (2003A0075-NS1-np and 2003B0114-NSb-np).
This work was partly supported by a Grant-in-Aid for Scientific Research (C) (Grant No. 15540329) from the Ministry of Education, Science, Sports, and Culture of Japan.


\begin{thebibliography}{}
\bibitem{SC} The metallic region should show superconductivity at 4 K, which is suggested by the bulk magnetization measurements \cite{Yoneyama1}.  We cite, however, a metal instead of a superconductivity in the higher frequency region at 4 K because it is difficult to distinguish between two states in the present experimental method.
\bibitem{beam} No picture elements are caused by an accidental interrupt of SR beam due to the unscheduled trouble of the storage ring.
\end{thebibliography}

\end{document}